\documentclass[aps,pra,twocolumn,showpacs,floatfix]{revtex4-1}
\usepackage[charter]{mathdesign}
\usepackage{amsmath}
\usepackage{graphicx}
\usepackage{bm}
\usepackage{color}
\usepackage{dcolumn}
\usepackage{amsfonts}

\def\be{\begin{equation}}
\def\ee{\end{equation}}
\def\ber{\begin{eqnarray}}
\def\eer{\end{eqnarray}}
\def\nn{\nonumber}

\def\qv{{ \bm q}}

\begin{document}

\title{Density-Wave Instability and Collective Modes in a Bilayer System of Dipolar Bosons}
\author{E. Akaturk}
\affiliation{Department of Physics, Bilkent University, 06800 Ankara, Turkey}
\author{S. H. Abedinpour}
\email{abedinpour@iasbs.ac.ir}
\affiliation{Department of Physics, Institute for Advanced Studies in Basic Sciences (IASBS), Zanjan 45137-66731, Iran}
\affiliation{School of Physics, Institute for Research in Fundamental Sciences (IPM), Tehran 19395-5531, Iran}
\author{B. Tanatar}
\affiliation{Department of Physics, Bilkent University, 06800 Ankara, Turkey}
\date{\today}

\begin{abstract}

We consider a bilayer of dipolar bosons in which the polarization of dipoles are perpendicular to the planes.
Using accurate static structure factor $S(q)$ data from hypernetted-chain calculation for single layer dipolar bosons we construct effective screened interactions for intralayer particles. We adopt the random-phase approximation for interlayer interactions.
We study the instability of the homogeneous bilayer system against the formation of density waves by investigating the poles of the density-density response function.
The dispersion of collective modes of this system also signals the density-wave instability.
We also investigate the effect of counterflow on the collective mode dispersion and on the density-wave instability 
and discuss the dissipationless superfluid drag effect in the presence of a background velocity.
\end{abstract}

\maketitle

\section{Introduction}\label{sec:intro}

Experimental progresses in trapping and cooling atoms with large magnetic moments and polar molecules, opened up a new and interesting area of exploring quantum many-body systems with large and anisotropic dipole-dipole interactions~\cite{baranov_phyrep08, lahaye_rpp09, baranov_chemrev2012}. 
Bose-Einstein condensation (BEC) of  polar molecules~\cite{ni_science2008,ni_physchem2009,ni_nature2010,deiglamyr_prl2008}, and atoms with large permanent magnetic moments~\cite{lahaye_nature2007,liu_prl2011,aikawa_prl2012} has been observed. 
Dipolar fermionic gases have been cooled down near to their ground-state as well~\cite{lu_prl2012, park_prl2015}.

In bulk geometries, the attractive part of the dipole-dipole interaction could in principle lead to instabilities, like BEC collapse~\cite{koch_nphys2008} or chemical reactions between particles~\cite{baranov_chemrev2012}. Therefore, it is usually favorable to confine the dipolar gases into quasi-two or one-dimensional geometries, and use an external electric or magnetic field (depending on the nature of dipoles) to polarize all dipoles in the same direction. 
Layered structures are another configuration of great interest which one can tune the attractive interactions and pairing between different layers without the fear of having chemical reactions.

While the stripe or density-wave phase is naturally expected in an isolated 2D system of tilted dipolar bosons~\cite{macia_prl2012,macia_pra2014} and fermions~\cite{yamaguchi_pra2010,sun_prb2010,parish_prl2012} due to the anisotropy of the dipole-dipole interaction, this instability has been the subject of much dispute in the limit of perpendicular dipoles, where the inter-particle interaction is isotropic.
While mean-field approximation~\cite{yamaguchi_pra2010} as well as density-functional theory (DFT)~\cite{vanzyl_arxiv2015} and Singwi-Tosi-Land-Sj{\"o}lander (STLS)~\cite{parish_prl2012} calculations all predict stripe phase formation at relatively low interaction strength for two-dimensional dipolar fermions, quantum Monte Carlo (QMC) simulations find that the stripe phase never becomes energetically favorable, up to the liquid-to-solid phase transition for perpendicular bosons~\cite{macia_pra2014} and fermions~\cite{matveeva_prl_2012}.

In double-layer structures, the fermionic systems attracted enormous attention. For example the ground state properties and instabilities have been studied within the Hartree-Fock~\cite{zinner_epjd2011,block_njp2012,babadi_prb2011} as well as STLS methods~\cite{marchetti_prb2013}. The QMC simulation has been employed to study the crossover from BEC to Bardeen-Cooper-Schrieffer (BCS) state~\cite{matveeva_pra2014} too.
Bosonic bilayers, on the other hand, received slightly less attention. Hufnagl and Zillich~\cite{hufnagl_pra2013} used  hypernetted-chain approximation (HNC) to calculate the ground-state quantities of a bilayer system of \textit{tilted} dipolar bosons. Then using the correlated-basis function method they obtained the dynamical properties.
More recently, the competition between single-dipole and dimer condensation in a bilayer of perpendicular dipolar bosons have been investigated using QMC method by Macia \textit{et al.}~\cite{macia_pra2014_2}
They have observed that at strong interlayer coupling, the pair superfluidity dominates over the single-particle superfluidity.    

In this work we study two parallel layers of dipolar bosons, whose moments are aligned perpendicular to the 2D-plane. 
Therefore, both the intralayer and interlayer interactions are isotropic. The bare intralayer interaction is purely repulsive while the bare interlayer interaction could be either repulsive or attractive, depending on the in-plane separation of two dipoles [see, Eqs.\eqref{eq:vsr} and \eqref{eq:vdr}, below].
We investigate the possibility of instability of a homogeneous liquid towards the formation of inhomogeneous densities, \textit{i.e.} density waves. For this purpose we look at the poles of the static density-density response function. The effective intralayer interactions are obtained from an accurate HNC results for the static structure factor of an isolated 2D layer~\cite{abedinpour_pra2012}, combined with the fluctuation-dissipation theorem~\cite{abedinpour_annphys2014}, and the interlayer interactions are treated within the random-phase approximation (RPA)~\cite{Giuliani_and_Vignale}. We also find the full dispersion of in-phase and out-of-phase collective modes (\textit{i.e.} zero-sound modes) from the poles of the dynamical density-density response function. We have also included a finite counterflow between two layers and studied its effects on the density-wave instability and collective modes. 
We finally address the dissipationless superfluid drag in the presence of a finite background flow.

A similar study of the instability of a homogeneous liquid with respect to the inhomogeneous phase of charge density wave has been studied in a variety of quantum charged systems ranging from single-layer electron gas~\cite{Sander} to electron-electron and electron-hole double-layers~\cite{Szymanski94,Swierkowski96,Neilson93,Moudgil02} to charged Bose systems~\cite{Moudgil97} and superlattices~\cite{Gold}.

The rest of this paper is organized as follows.
In section~\ref{sec:effective}, we introduce the density-density response function of our system and explain how effective intralayer interaction could be obtained from the static structure factor. 
In section~\ref{sec:instability}, we calculate the density-wave instability and the dispersion of the collective modes.  
Finally in section~\ref{sec:conclusion}, we summarize and conclude our main findings.

\section{Density-density response function and the effective interactions}\label{sec:effective}

We consider two identical two-dimensional planes of bosonic dipoles, separated by distance $d$. All dipoles are assumed to be polarized perpendicular to the planes. The bare intralayer and interlayer interactions respectively read
\be\label{eq:vsr}
V_s(r)=\frac{C_{\rm dd}}{4\pi}\frac{1}{ r^3}~,
\ee
and
\be\label{eq:vdr}
V_d(r)=\frac{C_{\rm dd}}{4\pi } \frac{r^2-2 d^2}{(r^2+d^2)^{5/2}}~,
\ee
where $C_{\rm dd}$ is the dipole-dipole coupling constant and $r$ is the in-plane distance between two dipoles.
After Fourier transformation one finds~\cite{block_njp2012}
\be\label{eq:vs}
V_s(q)=\frac{C_{\rm dd}}{4}\left[\frac{8}{3\sqrt{2\pi} w}-2 q e^{q^2 w^2/2}{\rm erfc}\left(\frac{qw}{\sqrt{2}}\right)\right]~,
\ee
and
\be\label{eq:vd}
V_d(q)=-\frac{C_{\rm dd}}{2}q e^{-q d}~.
\ee
Here erfc($x$) is the complementary error function and $w$ is the short distance cut off introduced to heal the divergence of Fourier transform of the intralayer interaction.

In this work we are interested in the density-wave instabilities and collective density modes of this double layer structure.  For this we begin with the following matrix equation for the density fluctuations~\cite{Giuliani_and_Vignale}
\be
\delta n_i(\qv,\omega)= \sum_j \chi_{ij}(\qv,\omega)  V^{\rm ext}_j (\qv,\omega)~,
\ee
where $\delta  n_i(\qv,\omega)$ is the density fluctuation in layer $i$ ($i=1,2$). $V^{\rm ext}_j(\qv,\omega)$ is the external potential applied to layer $j$ and $\chi_{ij}(\qv,\omega)$ is the density-density response function, whose matrix form reads
\be\label{eq:chi-1}
{\hat \chi}^{-1} (\qv,\omega)={\hat \Pi}^{-1}(\qv,\omega)- \hat{W}^{\rm eff}(\qv,\omega)~.
\ee
Here $\Pi_{ij}(\qv,\omega)=\delta_{ij}\Pi_i(\qv,\omega)$ is the non-interacting density-density response function of layer $i$, and $ W^{\rm eff}_{ij}(\qv,\omega)$ is the dynamical effective potential. 
For symmetric bilayers we have $\Pi_i(\qv,\omega)=\Pi(\qv,\omega)$ (same for both layers),
and $W^{\rm eff}_{ij}(\qv,\omega)=\delta_{ij}W_s(\qv,\omega)+(1-\delta_{ij})W_d(\qv,\omega)$, where $W_s(\qv,\omega)$ [$W_d(\qv,\omega)$] is the effective interaction between dipoles in the same [different] layers. 

Eigenvalues of the density-density response matrix ${\hat \chi}(\qv,\omega)$ are
\be\label{eq:chi_pm}
\chi_\pm(\qv,\omega)= \frac{\Pi(\qv,\omega)}{1-\Pi(\qv,\omega) W_\pm(\qv,\omega)}~,
\ee
where $W_\pm(\qv,\omega)=W_s(\qv,\omega) \pm W_d(\qv,\omega)$ are the symmetric and antisymmetric components of the effective potentials.

The non-interacting density-density response function $\Pi(\qv,\omega)$ of a two dimensional system of bosons is analytically known at zero temperature
\be\label{eq:lindhard}
\Pi(q, \omega)=\frac{2 n\varepsilon_q}{(\hbar \omega+i 0^+)^2-\varepsilon_q^2}~,
\ee
where $n$ is the particle density in one layer and $\varepsilon_q=\hbar^2q^2/(2m)$ is the single-particle energy. The exact form of the effective potentials are not known, and one has to resort to some approximations. In the celebrated random phase approximation~\cite{Giuliani_and_Vignale}, the effective potentials are replaced with their bare values. 
In this work we will use the bare interlayer potential $V_d(q)$ from Eq.~(\ref{eq:vd}). On the other hand, the bare intralayer potential, in $q$-space,~(\ref{eq:vs})  has an artificial cut-off dependence. In order to overcome this problem, we will use the fluctuation-dissipation theorem to find an approximate expression for the interlayer potential~\cite{abedinpour_annphys2014}. At zero temperature the fluctuation-dissipation theorem reads~\cite{Giuliani_and_Vignale}
\be\label{eq:fdt}
S(q)=-\frac{\hbar\pi}{n}\int_0^\infty {\mathrm d}\omega\,\Im m\left[\frac{\Pi(q,\omega)}{1-W_s(q,\omega)\Pi(q,\omega)}\right]~.
\ee
Now, neglecting the frequency dependence of the effective potential \emph{i.e.} replacing $W_s(q,\omega)$ with a static and real function $W_s(q)$, we can perform the frequency integral in Eq.~(\ref{eq:fdt}) analytically and obtain
\be\label{eq:ws}
W_s(q)=\frac{\varepsilon_q}{2 n}\left[\frac{1}{S^2(q)}-1\right]~.
\ee
Here, $S(q)$ is the static structure factor of an isolated 2D dipolar Bose liquid, which can be obtained very accurately {\it e.g.}, from QMC simulations~\cite{astrakharchik_prl2007,Buchler} or HNC calculations~\cite{abedinpour_pra2012}. The effects of exchange and correlation, which are entirely ignored in the RPA, are partly included in Eq.~(\ref{eq:ws}) through the static structure factor. 
We set the interlayer part of the effective interaction to the bare interlayer interaction
i.e., $W_d(q)=V_d(q)$ since a QMC or HNC calculation for interlayer static structure factor is not available. Such an approximation is equivalent to RPA and we surmise
it will be adequate for large enough layer separations.  

\section{Density-Wave Instabilities and Collective Modes}\label{sec:instability}
Density-wave instabilities and collective modes could be obtained from the poles of the density-density response function 
given in Eq.\,(\ref{eq:chi_pm}), respectively in the  static and dynamical limits, or equivalently from the solution of
\be\label{eq:dielectric}
1-\Pi(\qv,\omega)W_\pm(q,\omega)=0~,
\ee

\begin{figure}
\begin{tabular} {c}
    \includegraphics[scale=0.33]{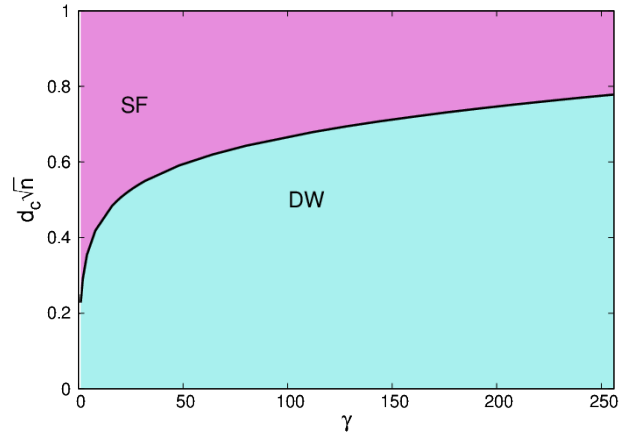}
   \end{tabular}
       \caption{The critical layer separation $d_{c}$ 
       versus coupling constant $\gamma$ for bilayer dipolar bosons.
       The  purple region shows the homogeneous superfluid (SF) phase and the green one is the region with density-wave (DW) instability.
\label{fig:qc_dc}} 
\end{figure}

\emph{Density-wave instability:} 
In the static limit, the non-interacting density-density response of Eq.~(\ref{eq:lindhard}) reduces to
\be\label{eq:lind_static}
\Pi(\qv)=-\frac{2 n}{\varepsilon_q}~,
\ee
which together with Eq.~(\ref{eq:dielectric}), gives
\be\label{eq:dwi}
1+\frac{2 n}{\varepsilon_\qv}W_\pm(q) =0~.
\ee
Now, using the bare interlayer potential~\eqref{eq:vd} and the effective intralayer potential of \eqref{eq:ws} in Eq.~\eqref{eq:dwi} we find
\be\label{eq:qc}
q= \pm 8 \pi n r_0 S^2(q)  e^{-q d}~.\nn
\ee
Here 
$r_0=m C_{\rm dd}/(4\pi\hbar^2)$ is a characteristic length scale. 
As the static structure factor is positive by definition, the above expression with minus sign will not have any solution which means that no density-wave singularity in the \emph{out-of-phase} channel (\emph{i.e.}, $\chi_-=0$) appears. On the other hand, in the \emph{in-phase} channel (\emph{i.e.}, $\chi_+=0$) one can find instabilities for suitable values of the interaction strength and layer spacing from the solutions of Eq.~(\ref{eq:qc}) with the positive sign. 
Numerical investigation of Eq.~(\ref{eq:qc}) reveals that (see, Fig.\,\ref{fig:qc_dc}) at every value of the coupling constant $\gamma=n r_0^2$, there is a critical layer spacing $d_c$, below which one can find density-wave instability at a finite value of the wave vector $q_c$. 

We note that for an isolated single layer, one has $W_d(q)=0$, and the criteria for the density-wave instability becomes
\be
\begin{split}
0=1-W_s(q)\Pi(q)
=\frac{1}{S^2(q)}~,
\end{split}
\ee
which has no solution at any finite $q$. Therefore, within the approximations  we use here, no density-wave instability is expected to happen in an isolated two-dimensional system with purely repulsive dipolar interaction.
In agreement with the QMC findings~\cite{macia_pra2014}.

\begin{figure}
\begin{tabular} {c}
    \includegraphics[width=0.4\textwidth]{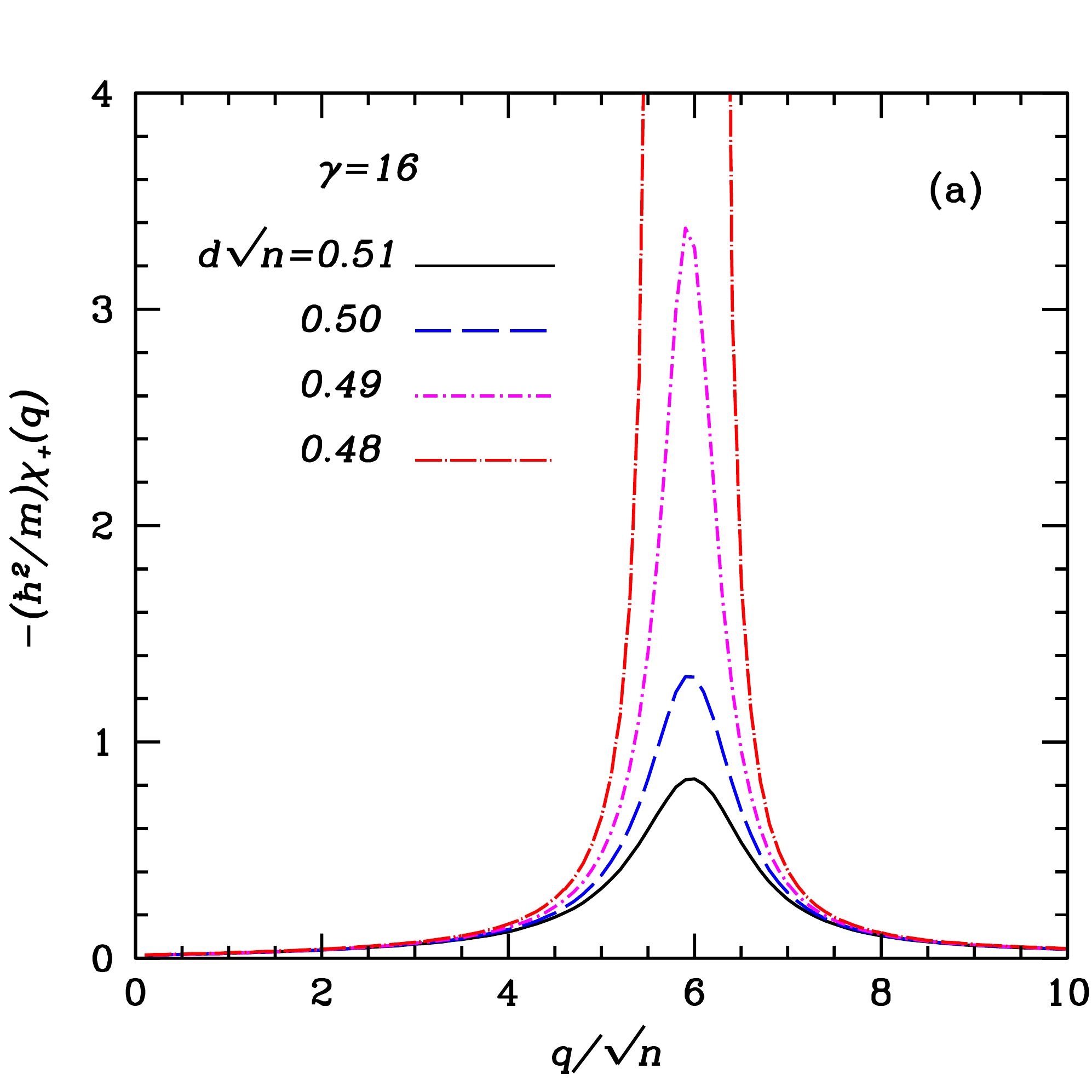}\\
    \includegraphics[width=0.4\textwidth]{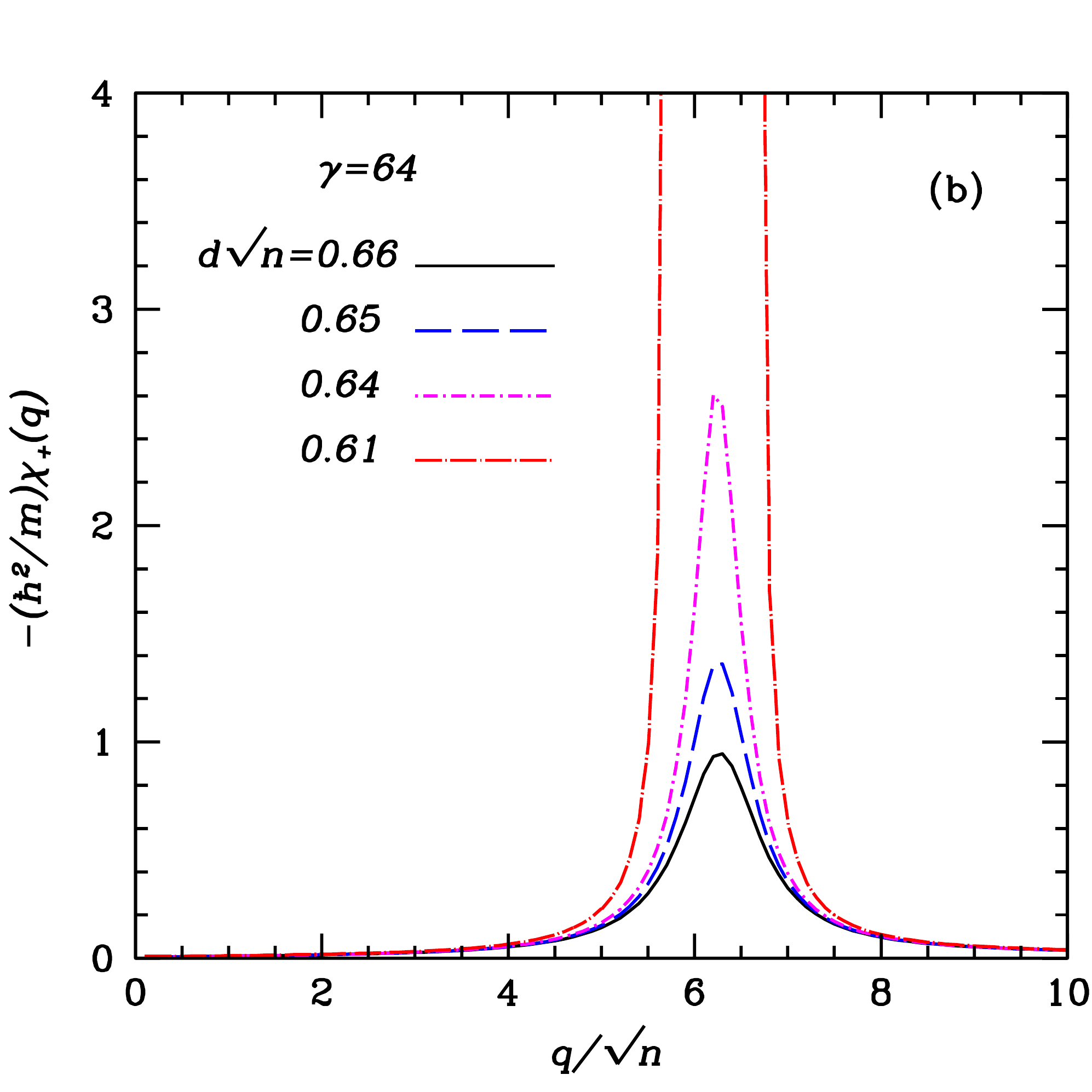}
 \end{tabular}
   \caption{The symmetric component of the static density-density response function $\chi_+(q)$ as a function of the dimensionless wave vector $q /\sqrt{n}$ for several values of the layer spacing $d$. As $d$ approaches the critical spacing $d_c$, a singularity at finite $q$ emerges in the density-density response function. Top and bottom panels are for two different values of the coupling constant, $\gamma=16$ and $\gamma=64$, respectively.  \label{fig:chi_p}}  
\end{figure}

\begin{figure}
\begin{tabular} {c}
    \includegraphics[width=0.4\textwidth]{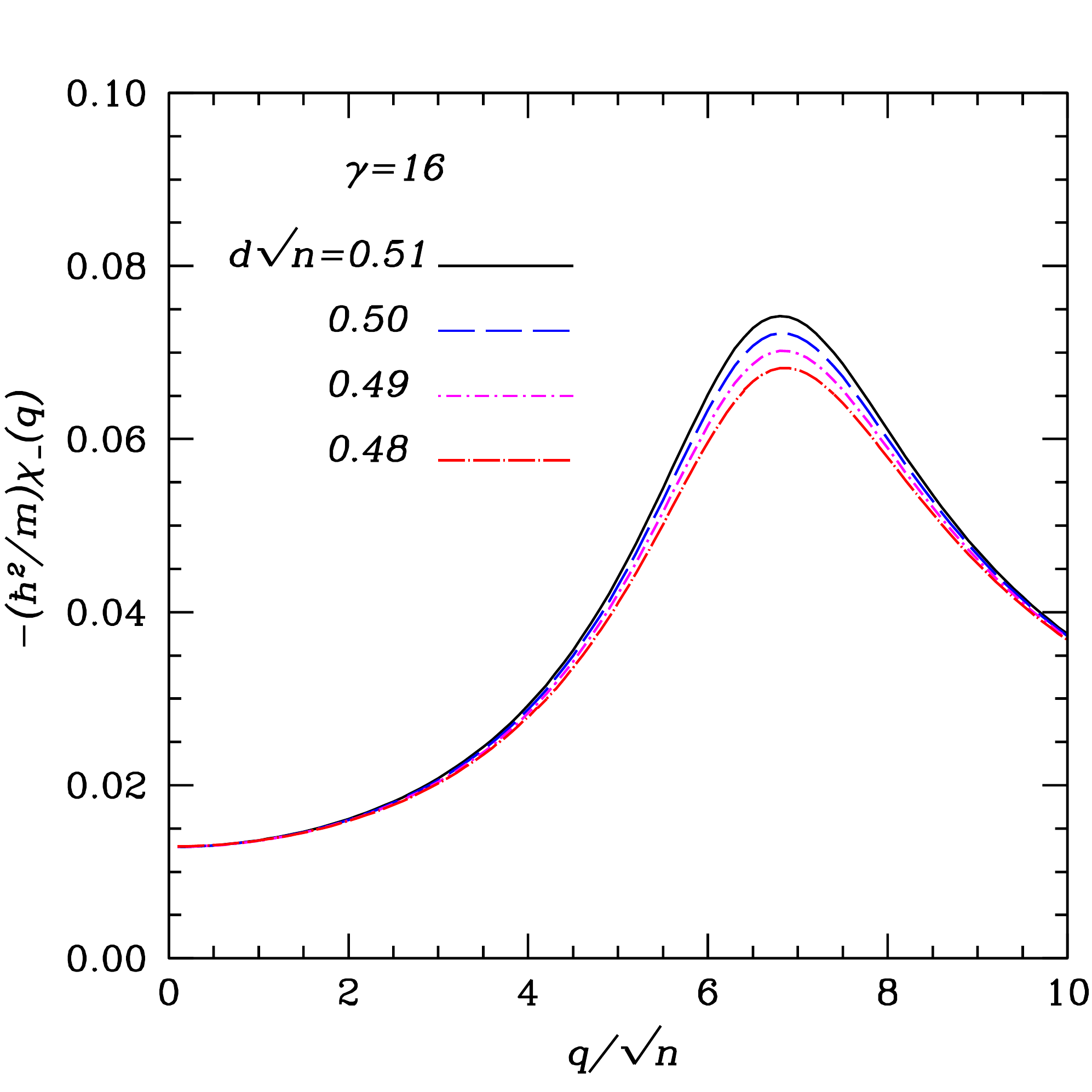}
 \end{tabular}
   \caption{The antisymmetric component of the static density-density response function $\chi_-(q)$ as a function of the dimensionless wave vector $q /\sqrt{n}$ for several values of the layer spacing $d$ at $\gamma=16$.  
 \label{fig:chi_m}}  
\end{figure}

The behavior of static density-density response functions
\be
\chi_{\pm}(q)=-\frac{4 m n}{\hbar^2}\frac{1}{\frac{q^2}{S^2(q)}\mp 8\pi n r_0 q e^{-q d}}~,
\ee
are also illustrated in Figs.~\ref{fig:chi_p} and~\ref{fig:chi_m}. As the layer separation is lowered to the critical distance, a singularity in $\chi_+(q)$ emerges (see, Fig.~\ref{fig:chi_p}), but the antisymmetric component of the density-density response function $\chi_{-}(q)$, remains finite.

\begin{figure}
\begin{tabular} {c}
    \includegraphics[width=0.4\textwidth]{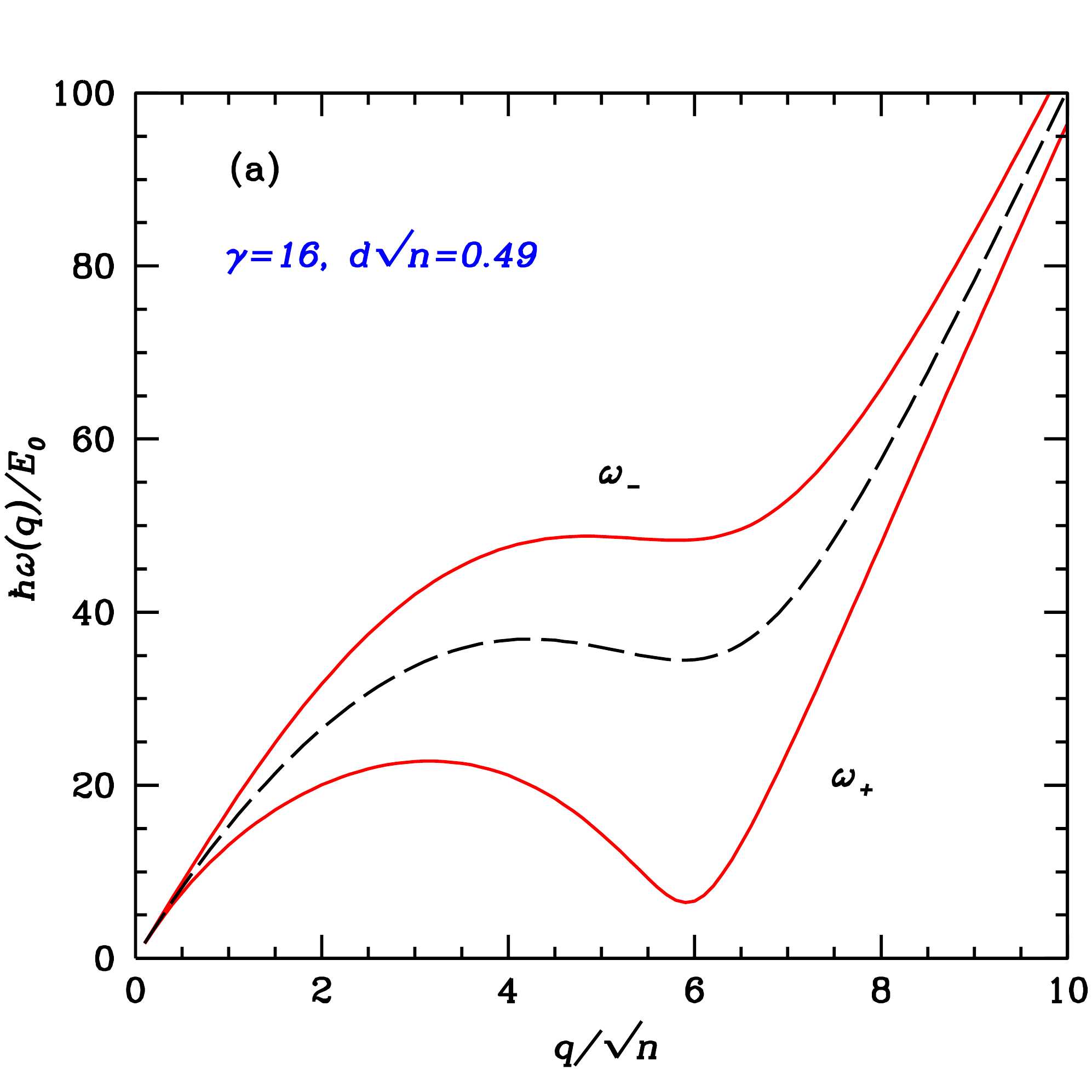}\\
        \includegraphics[width=0.4\textwidth]{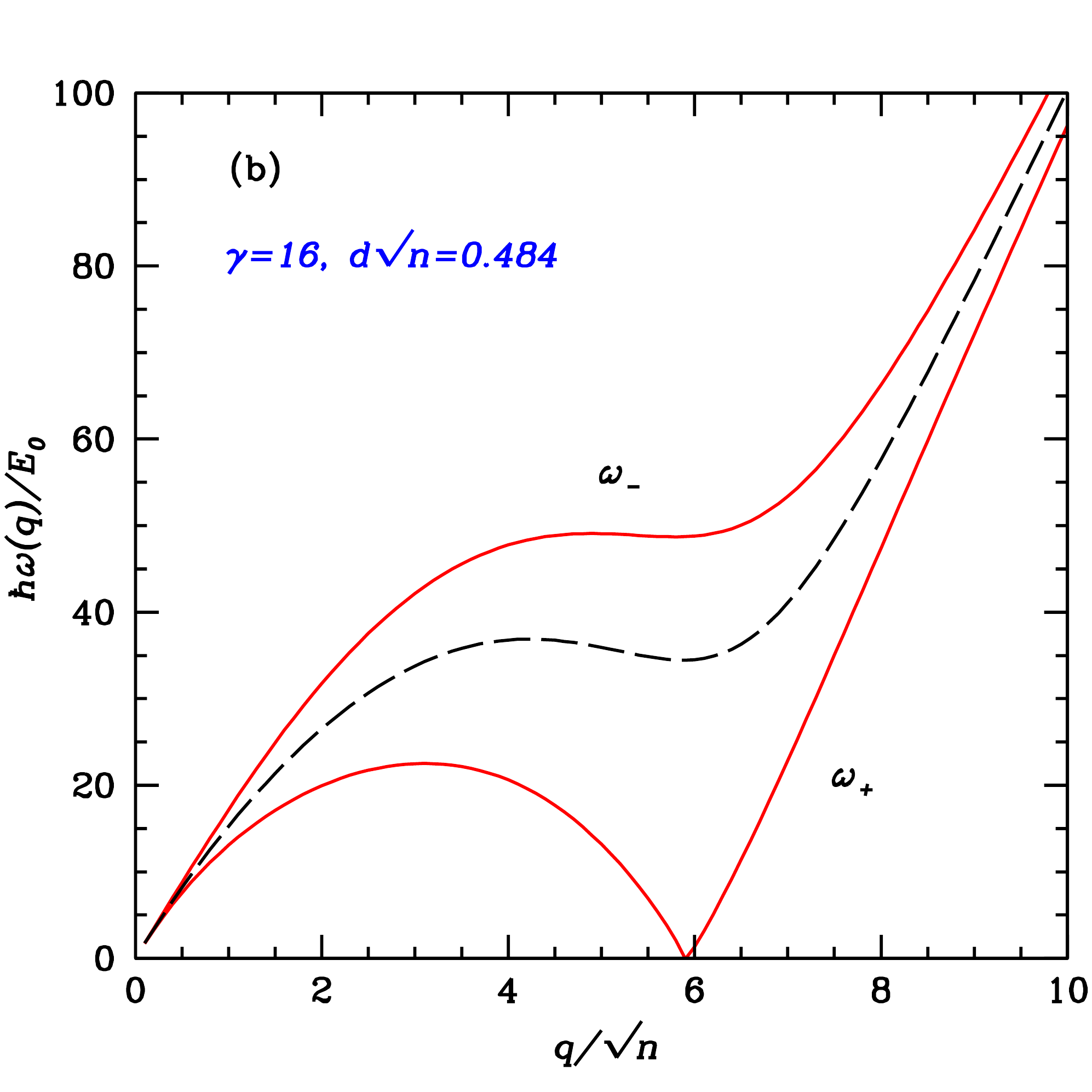}\\
    \includegraphics[width=0.4\textwidth]{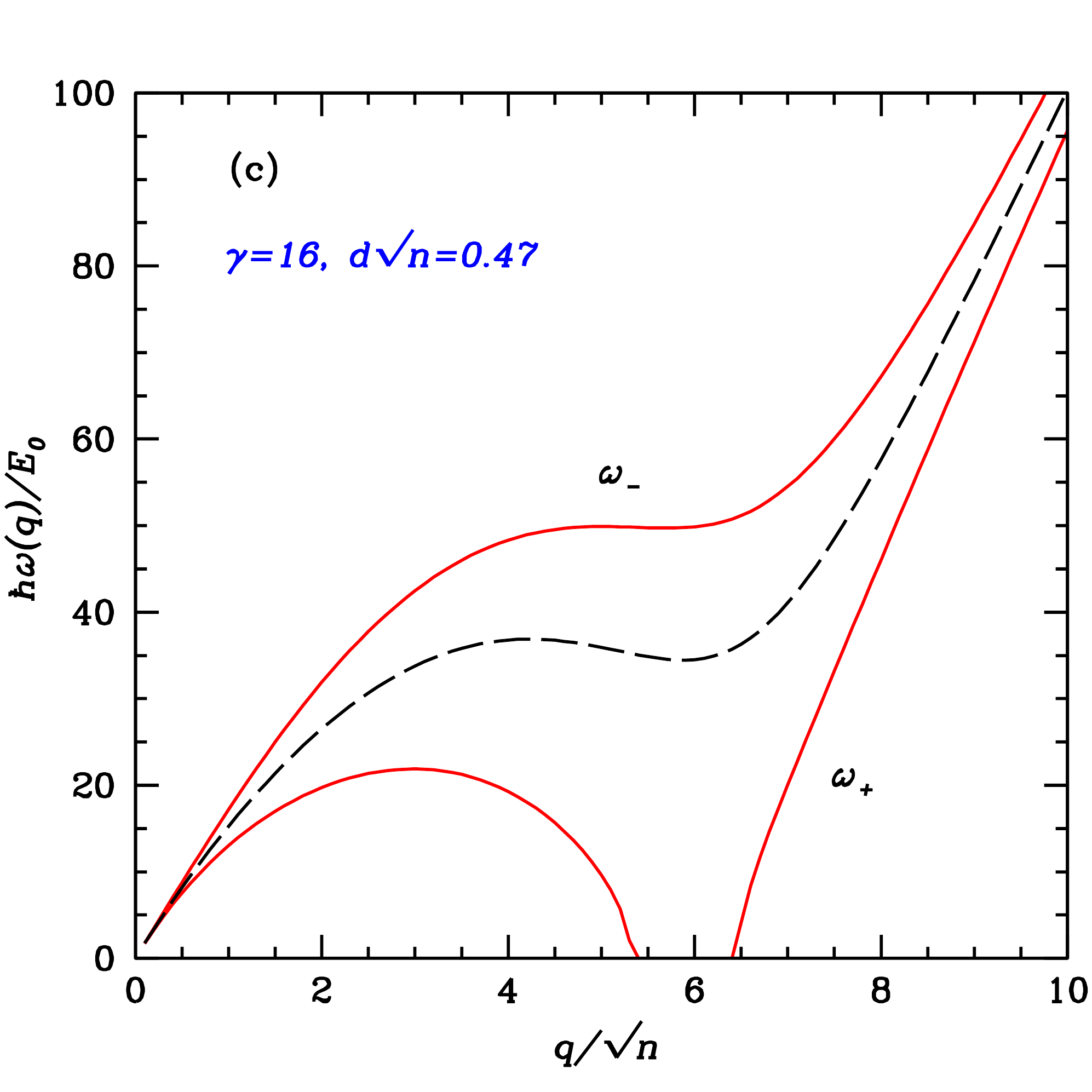}
   \end{tabular}
       \caption{Dispersion of symmetric $\omega_+$ and antisymmetric $\omega_-$ modes at a fixed value of the coupling constant $\gamma=16$, and for different values of the layer separation: $d=0.49$ (top), $d=0.484$ (middle), and $d=0.47$ (bottom). The dashed line represents single-layer collective mode $\omega_{sl}(q)$. Note that $d=0.484$ is the critical value of the layer separation of the formation of density-wave instability for $\gamma=16$.
\label{fig:wpm_g}} 
\end{figure}

\emph{Collective modes:} 
In order to find the collective modes, we study the singularities of the dynamical response functions $\chi_\pm(q,\omega)$. If we approximate the dynamical effective interactions with static and real functions, we find from Eq.~(\ref{eq:dielectric}):
\be
\hbar^2 \omega^2_\pm(q)=\varepsilon^2_q + 2n \varepsilon_q W_\pm(q)~.
\ee
Again, replacing the effective interlayer potential $W_d(q)$ with the bare interaction $V_d(q)$, and the effective intralayer potential $W_s(q)$ from Eq.~(\ref{eq:ws}), the dispersion of collective modes read
\be\label{eq:omega_pm}
\omega^2_\pm(q)=\frac{\varepsilon_q}{\hbar^2}\left[\frac{\varepsilon_q}{S^2(q)}\mp n C_{dd} q e^{-q d}~\right]~.
\ee
Note that the first term on the right-hand-side of this equation is the Bijl-Feynman excitation spectrum of a single layer dipolar Bose liquid~\cite{abedinpour_pra2012}. In the long wavelength limit, as the static structure factor vanishes linearly [$S(q\to 0) \propto q$], we find
\be
\omega_\pm(q)\approx v_s q+{\cal O}(q^2)~,
\ee
where $v_s$ is the sound velocity. 
Note that unlike the bilayer charged boson system~\cite{tanatar_prb96}, both collective modes of a bilayer system of dipolar bosons have acoustic nature and are degenerate at long wavelengths. 
Using the numerical results for the static structure factor from~\cite{abedinpour_pra2012} in Eq.~(\ref{eq:omega_pm}), the full dispersion of the collective modes could be readily obtained. The results for $\omega_\pm(q)$ and single-layer collective mode $\omega_{sl}(q)=\varepsilon_q/[\hbar S(q)]$ [in units of $E_0=\hbar^2 n/(2m)$] are presented in Fig.\,\ref{fig:wpm_g} for a fixed value of the coupling constant 
$\gamma$  and for different values of the layer separation $d$.  We find that the critical layer spacing is $d_c= 0.484/\sqrt{n}$ for $\gamma=16$, at which point the
symmetric mode $\omega_+$ touches zero and becomes soft. This occurs at the same $q$-value that the symmetric density-density response function $\chi_+$ diverges (c.f. Fig.\,\ref{fig:chi_p}).

\begin{figure}
\begin{tabular} {c}
    \includegraphics[width=0.4\textwidth]{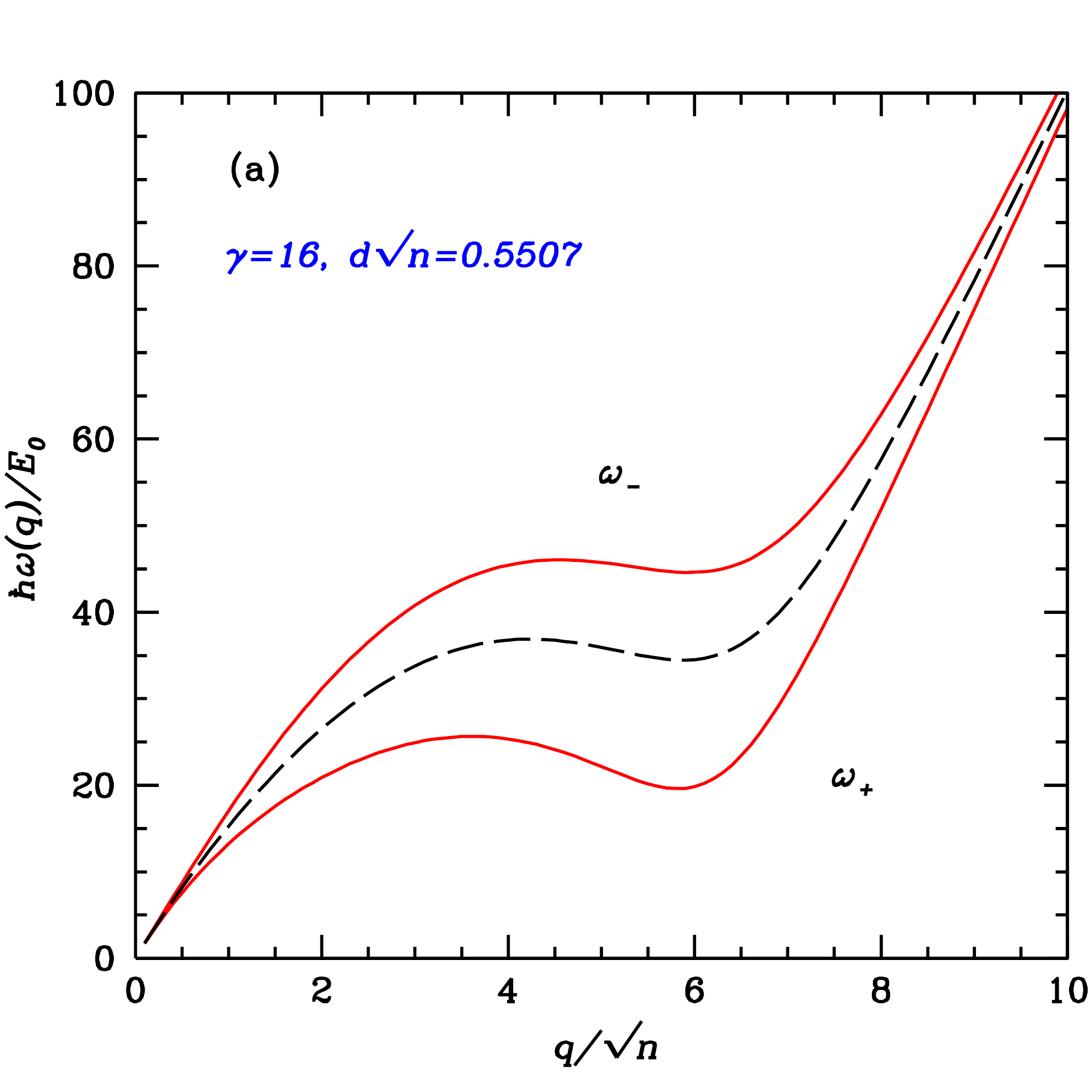}\\
     \includegraphics[width=0.4\textwidth]{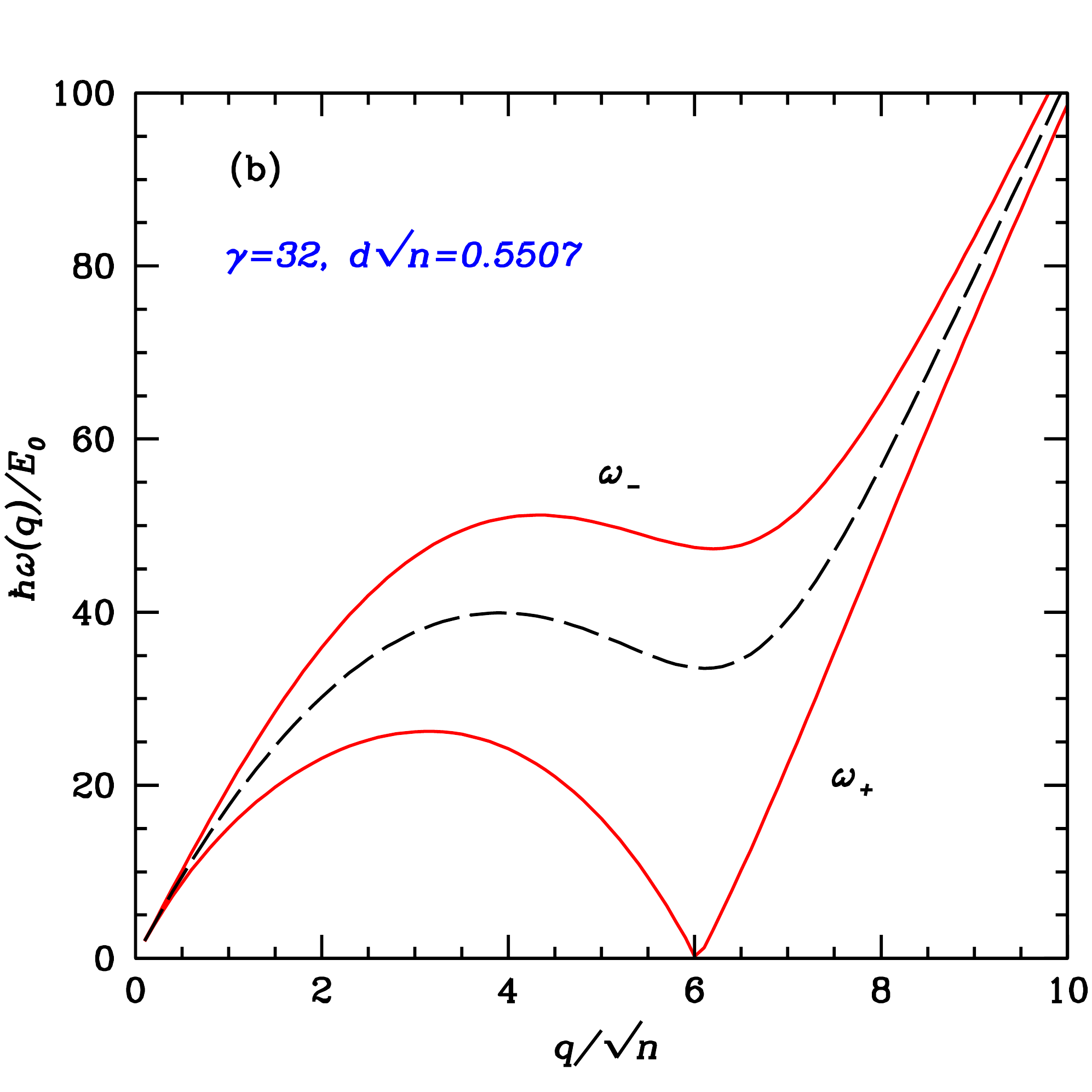}\\
      \includegraphics[width=0.4\textwidth]{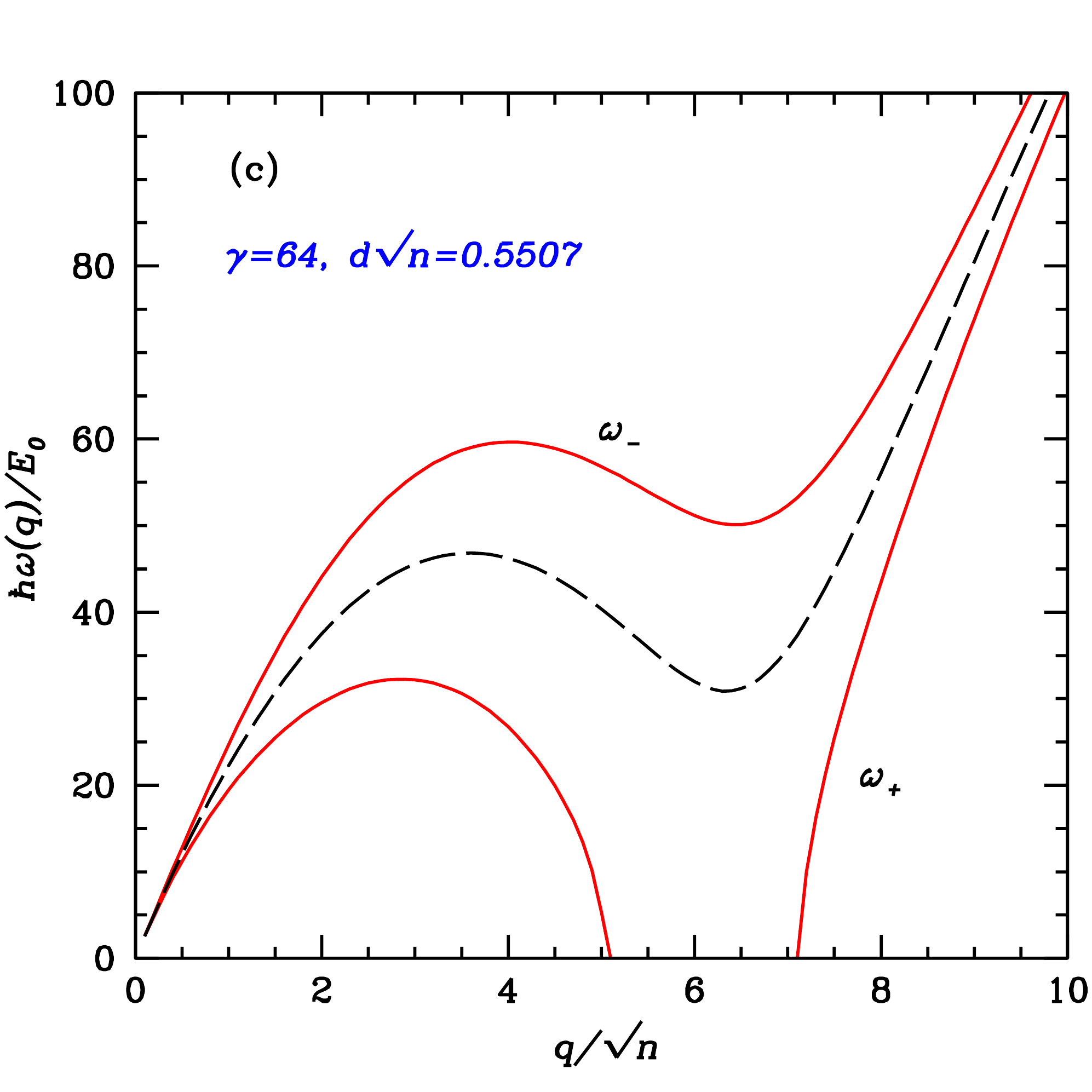}
   \end{tabular}
       \caption{Dispersion of collective modes of a bilayer system of dipolar bosons for different values of the coupling constant $\gamma=16$ (top), $\gamma=32$ (middle), and $\gamma=64$ (bottom) and for a fixed value of the layer spacing $d=0.5507$. The dashed line represents single-layer collective mode $\omega_{sl}(q)$. 
          \label{fig:wpm_dc}} 
\end{figure}

In Fig.\,\ref{fig:wpm_dc} the full dispersions of collective modes are presented at
different values of the coupling constant $\gamma$. We kept the layer spacing at  the critical value of the layer $d_c=0.5507/\sqrt{n}$ for $\gamma=32$ and observed a similar mode softening with increasing $\gamma$.

\emph{Counterflow:}
Now we consider background velocities of ${\bf v}_1$ and ${\bf v}_2$ in the first and second layers respectively. The collective modes and density-wave instabilities in the presence of these background velocities could be easily obtained from the poles of the total density-density response, after replacing $\omega$ with  $\omega-{\bf v}_i\cdot \qv$ in the noninteracting density response of layer i, $\Pi_i(q,\omega)$. The background flows could be decoupled into the center-of-mass ${\bf V}=({\bf v}_1+{\bf v}_2)/2$ and counterflow ${\bf v}=({\bf v}_1-{\bf v}_2)/2$ components. As the effect of the center-of-mass flow could be simply understood in terms of a Galilean boost, we focus on the counterflow part, and look for the solutions of the following equation
\be
\left|
\begin{array}{cc}
\dfrac{1}{\Pi(q,\omega-{\bf v}\cdot \qv)}-W_s(q, \omega)& -W_d(q, \omega) \\
-W_d(q, \omega) & \dfrac{1}{\Pi(q,\omega + {\bf v}\cdot \qv)}-W_s(q, \omega)
\end{array}
\right|=0~.
\ee
In the  static limit ($\omega=0$), we find
\be
1+\frac{2 n}{\varepsilon_q}W_{\pm}(q)-\left(\frac{\hbar {\bf v}\cdot \qv}{\varepsilon_q}\right)^2=0~.
\ee
Now, using the same approximation for the intra- and inter-layer potentials as before, we arrive at
\be\label{eq:qc_v}
q=\pm 8\pi n r_0 S^2( q) e^{-d q}+\frac{4 m^2 v^2 \cos^2{\theta}}{\hbar^2 q}S^2(q)~,
\ee
where $\theta$ is the angle between counterflow direction and the $q$ vector. 
Eq.~(\ref{eq:qc_v}) clearly shows that at a finite counterflow, the density-wave instability in the symmetric channel is facilitated in the directions parallel to the flow.

Similarly, for the collective modes in the presence of a counterflow, we obtain
\be
\begin{split}
&\hbar^2\omega^2_\pm(\qv,v)=\varepsilon^2_q\left[1+\frac{2 n}{ \varepsilon_q }W_s(q)\right]+\left(\hbar {\bf v}\cdot \qv\right)^2\\ 
&\mp 2 \varepsilon_q \sqrt{n^2 W^2_d(q)+\left(\hbar {\bf v}\cdot \qv\right)^2\left[1+\frac{2 n}{ \varepsilon_q }W_s(q)\right]}~.
\end{split}
\ee 
We note that the dispersions of collective modes become anisotropic in the presence of a finite counterflow.
Finally, replacing the effective interaction $W_s(q)$ from Eq.~(\ref{eq:ws}), and the bare interlayer interaction $V_d(q)$ from Eq.~(\ref{eq:vd}), we find
\be\label{eq:wpm_v_full}
\begin{split}
\hbar^2\omega^2_\pm(\qv,v)&=\left[\frac{\varepsilon_q}{S(q)}\right]^2+\left(\hbar {\bf v}\cdot \qv\right)^2\\
&\mp  \varepsilon_q \sqrt{ n^2 C^2_{dd} q^2 e^{-2 q d} +\left[\frac{2 \hbar {\bf v}\cdot \qv}{S(q)}\right]^2}~.
\end{split}
\ee
To leading order in the counterflow velocity $v$, we find for the dispersions 
\be\label{eq:wpm_v}
\omega_\pm(\qv,v) \approx \omega_\pm(q,0)+\frac{m S(q) v^2\cos^2(\theta)}{\hbar \sqrt{1\mp \alpha(q)}}\left[1\mp \frac{2}{\alpha(q)}\right]
+{\cal O}( v^4)~,
\ee
where
\be
\alpha(q)=\frac{8\pi n r_0}{q} S^2(q) e^{-qd}~.
\ee

\begin{figure}
    \includegraphics[width=0.4\textwidth]{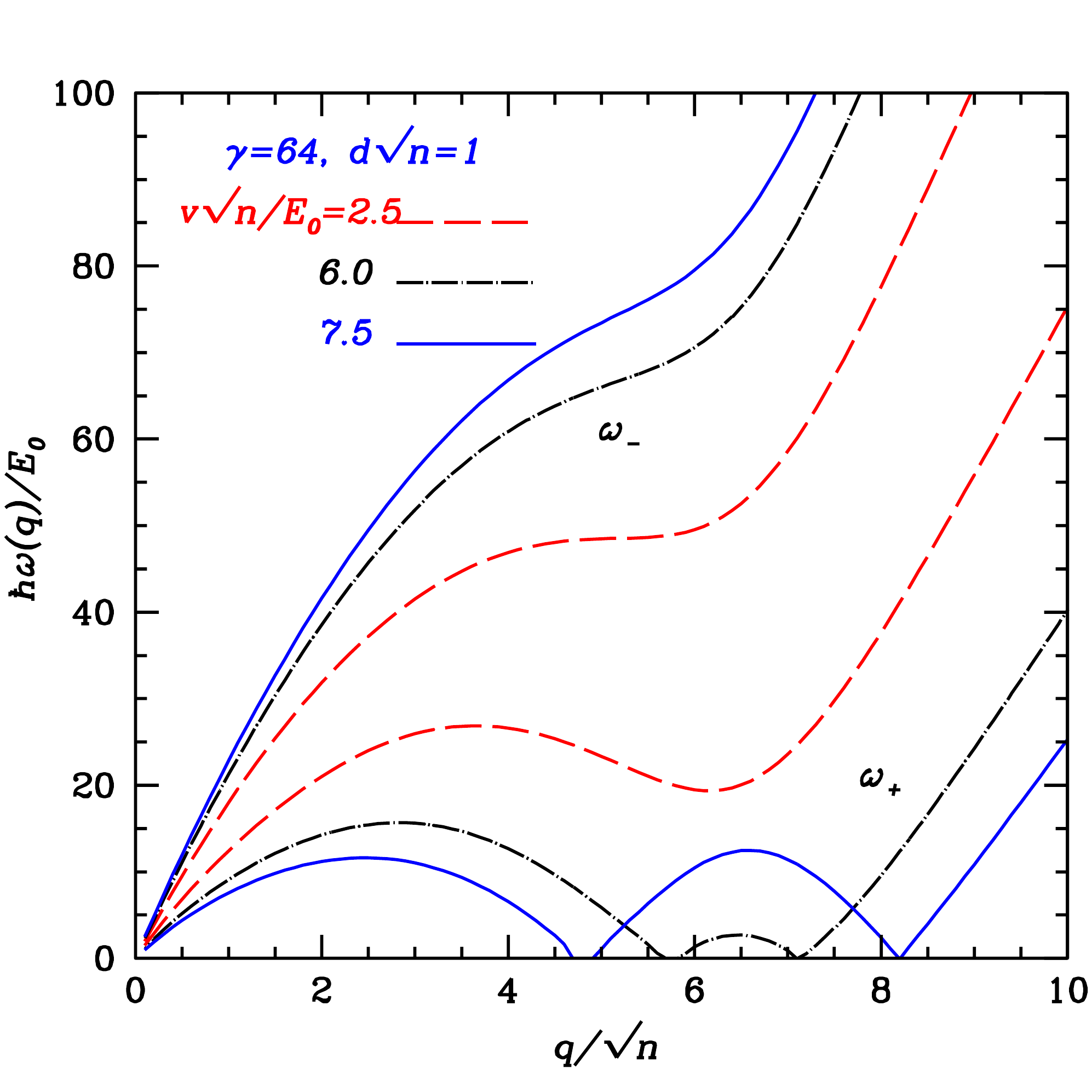}
       \caption{Dispersion of collective modes of a bilayer system of dipolar bosons at finite counterflow velocity $v$ and several values of the layer separation $d$. The interaction strength is $\gamma=16$. Note that we specialize to the case $\theta=0$ in Eq.\,(\ref{eq:wpm_v_full}), i.e., when the wave vector ${\bf q}$ is parallel to ${\bf v}$.
\label{fig:wpm_dc_v}} 
\end{figure}

In Fig.\,\ref{fig:wpm_dc_v} we illustrate the effect of counterflow velocity $v$ on the collective modes for a bilayer dipolar fluid at $\gamma=16$ and $d=1/\sqrt{n}$.
We note that the counterflow has the main effect of lowering the symmetric mode $\omega_+(q)$ and enhancing the antisymmetric mode $\omega_-(q)$. As $v$ increases symmetric mode becomes zero at two values of $q$ indicating instabilities according to Eq.\,(\ref{eq:qc_v}).

\begin{figure}
    \includegraphics[width=0.4\textwidth]{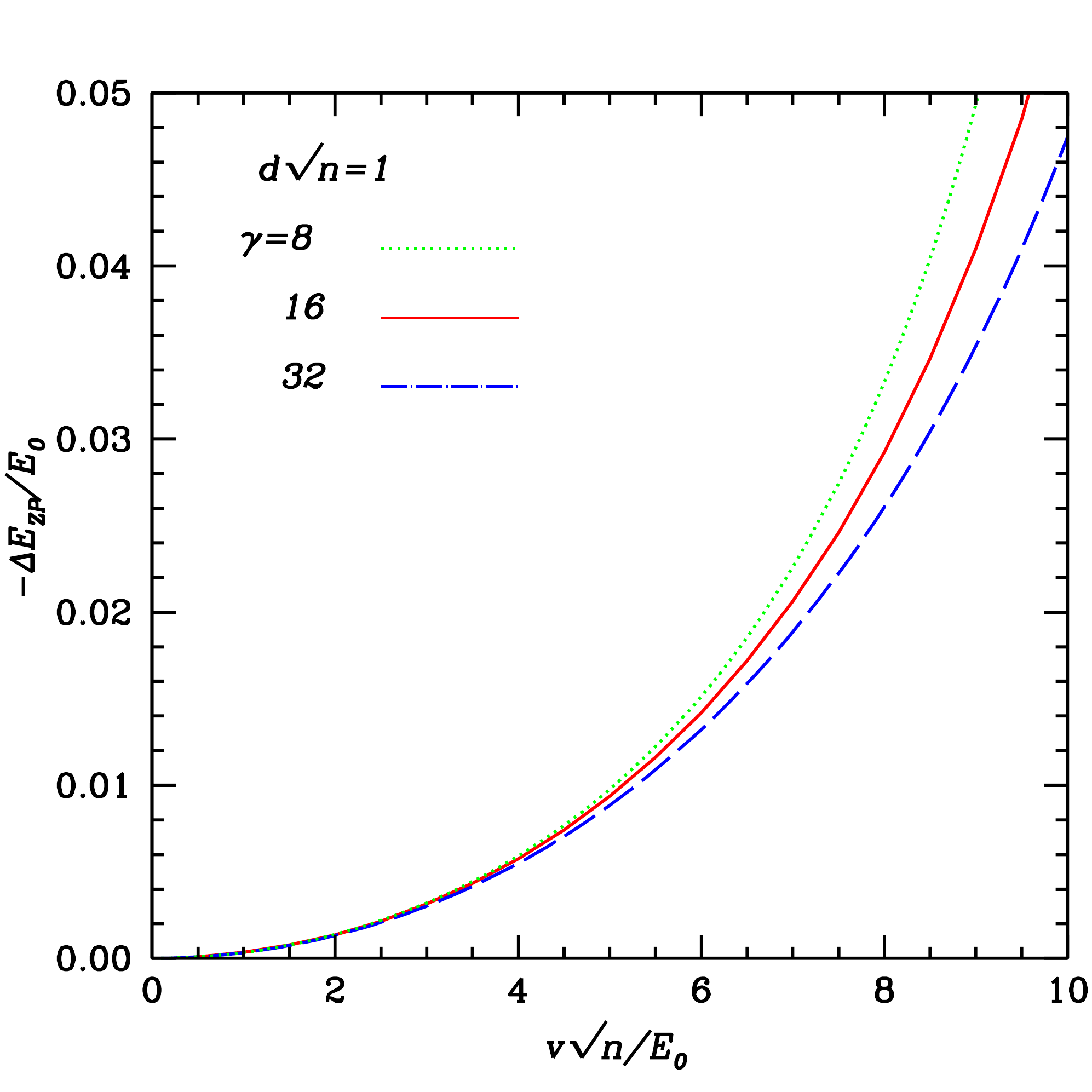}
       \caption{Change in the zero point energy $\Delta E_{\rm ZP}$ as a function of counterflow velocity $v$ and for several values of the interaction strength $\gamma$. The layer separation is $d=1/\sqrt{n}$.
 \label{fig:f7}} 
\end{figure}

\emph{Zero-point energy and drag effect:}
With the full dispersion relations, we are able to find the change in the zero-point energy, due to the finite counterflow \cite{duan_93}
\be
\Delta E_{\rm ZP}=\frac{\hbar}{2}\sum_{\qv,\alpha=\pm}  \left[ \omega_\alpha(q,v)-\omega_\alpha(q,0)\right]~,
\ee
in which the difference between collective modes with and without counterflow are summed.
Now, using the leading order results in $v$, from~(\ref{eq:wpm_v}) and assuming that the static structure factor is power expandable at small $q$, \emph{e.g.},
\be
S(q\to 0)=\frac{\hbar q}{2 m v_s}+ \beta_2 q^2 + \beta_3 q^3 +...~,
\ee
we find
\be
\Delta E_{\rm ZP} \approx-\frac{1}{A}\sum_{\qv}\left[ \frac{\hbar C^2_{dd} n^2 v^2 \cos^2(\theta)}{32 m^2 v_s^5} q^3+{\cal O}(q^4)\right]~.
\ee
Expansion of each dispersion alone depends on the higher order coefficients of $S(q)$, but interestingly they cancel out in the summation of two modes (up to $q^3$ terms).
Integrating this expression up to $q\sim 1/d$, we obtain
\be\label{eq:ezp_d}
\Delta E_{\rm ZP} \approx - \frac{\hbar C^2_{dd} n^2 v^2 }{640\pi m^2 v_s^5 d^5}.
\ee
We note that the zero-point energy is negative, meaning that a finite counterflow lowers the energy of the system, depends on the square of the velocity and scales as $d^{-5}$ with the separation of two layers. It is interesting to note that zero-point
energy for dipolar fermions yields $d^{-2}$ dependence on the separation 
distance. \cite{tanatar_jltp2013}
Previous calculations \cite{tanatar_prb96} on charged systems interacting via the
$1/r$ Coulomb potential have shown that the bilayer separation dependence of the zero-point energy is different for electrons and charged bosons too.  In Fig.\,\ref{fig:f7}
we show the numerical results of $\Delta E_{\rm ZP}$ as a function of the counterflow velocity $v$ at
$d=1/\sqrt{n}$ for various values of the coupling strength $\gamma$.

We now construct the free energy $F$, by adding the kinetic energies of the fermions in  each layer
\be\label{eq:fe}
F=\frac{1}{2}nm(v_1^2+v_2^2) - \frac{\hbar C^2_{dd} n^2}{2560 \pi m^2 v_s^5 d^5}(v_1-v_2)^2~,
\ee
where we have reverted to use the individual velocities in each layer. We find the current densities in layer 1 and layer 2, calculated from $j_i=\partial F/\partial v_i$
($i=1,2$), to be
\be\label{eq:j12}
\begin{array}{l}
j_1=\left(nm-\frac{\hbar C_dd^2n^2}{1280\pi m^2v_s^5d^5}\right)v_1
+\frac{\hbar C_dd^2n^2}{1280\pi m^2v_s^5d^5} v_2\, ,\\
\\
j_2=\frac{\hbar C_dd^2n^2}{1280\pi m^2v_s^5d^5}v_1
+\left(nm-\frac{\hbar C_dd^2n^2}{1280\pi m^2v_s^5d^5}\right)v_2\,
.
\end{array}
\ee
The expressions given in Eq.\,(\ref{eq:j12}) demonstrate that the superflow in the first (second) layer depends on the superfluid velocity on the same layer as well as that of the second (first) layer. This is the dissipationless superfluid drag effect well known in two-component superfluids~\cite{duan_93,tanatar_jltp2013,tanatar_prb96,shevchenko_97,andreev_75,alpar_84} which has been discussed for a variety of related systems.

\section{Summary and Conclusions}\label{sec:conclusion}
In summary, we have investigated the instability of a homogenous bilayer system of perpendicular dipolar bosons towards density waves. An accurate HNC results for the intralayer static structure factor is used together with the fluctuation-dissipation theorem to extract a static intralayer effective potential and the random phase approximation is employed for the interlayer interaction. We have observed that at any intralayer coupling strength, there is a critical layer spacing below which the homogenous in-plane density becomes unstable. 
The full dispersion of the in-phase and out-of-phase zero-sound modes of the bilayer system have been calculated too. We observed that two modes become degenerate in the long wave-length limit. 

Furthermore, we have studied the effects of a finite counterflow on the density-wave instability and collective modes. 
Counterflow lowers the zero-point energy and drive the homogenous system towards the formation of density waves.
We have also studied the dissipationless drag between the two layers of dipoles similar to other bilayer systems.

Finally, we should note that in the limit of closely separated layers, improvements beyond the RPA in the effective interlayer potential, like the inclusion of exchange-correlation effects or the binding of two dipoles from different layers might be necessary. Dynamical effects and frequency dependance of the effective potentials would become important in the strongly-correlated regime too.

\acknowledgments
This work is supported in part by TUBITAK Grant No. 112T974 and TUBA.
S.H.A. thanks the hospitality and support of Department of Physics at Bilkent University, during his visits.

\end{document}